\documentclass[12pt]{iopart}
\usepackage{iopams}
\usepackage{graphicx}
\usepackage{caption}
\usepackage{subcaption}
\usepackage{hyperref}
\usepackage{hypernat}
\usepackage{color}

\begin{document}

\title[Influence of cumulative damage on synchronization of Kuramoto oscillators]{Influence of cumulative damage on synchronization of Kuramoto oscillators on networks}
\author{L. K. Eraso-Hernandez${}^{1,*}$, A. P. Riascos$^{1,2}$}

\address{${}^1$Instituto de F\'isica, Universidad Nacional Aut\'onoma de M\'exico, 
Apartado Postal 20-364, 01000 Ciudad de M\'exico, M\'exico\\
${}^{2}$ Departamento de F\'isica, Universidad Nacional de Colombia, Bogotá, Colombia}
\ead{erasoleidy@estudiantes.fisica.unam.mx}

\begin{abstract}
In this paper, we study the synchronization of identical Kuramoto phase oscillators under cumulative stochastic damage to the edges of networks. We analyze the capacity of coupled oscillators to reach a coherent state from initial random phases. The process of synchronization is a global function performed by a system that gradually changes when the damage weakens individual connections of the network. We explore diverse structures  characterized by different topologies. Among these are deterministic networks as a wheel or the lattice formed by the movements of the knight on a chess board, and random networks generated with the Erd\H{o}s-R\'enyi and  Barab\'asi-Albert algorithms. In addition, we study the synchronization times of 109 non-isomorphic graphs with six nodes. The synchronization times and other introduced quantities are sensitive to the impact of damage, allowing us to measure the reduction of the capacity of synchronization and classify the effect of damage in the systems under study. This approach is general and paves the way for the exploration of the effect of damage accumulation in diverse dynamical processes in complex systems.
\end{abstract}

\maketitle
\section{Introduction}
Synchronization is a collective process in which a  coupled population, under certain conditions, becomes self-organized in such a way that their components evolve to follow the same dynamical pattern \cite{VespiBook,PikoBook,strogatzbook,J_Kurths_PhysRep2023}.  This process is one of the most attractive phenomena in nature and some common examples are the synchronization of flashing fireflies \cite{PikoBook,doi:10.1126/sciadv.abg9259}, a crowd clapping in unison \cite{neda}, synchronization in arrays of Josephson junctions \cite{josephson},  among many others \cite{PikoBook,strogatzbook,Balanov2009}. Synchronization process is considered universal \cite{PikoBook} so it is present in extremely diverse systems and plays a fundamental role in their functioning. In particular, synchronous activity is pivotal in groups of living beings such as colonies of insects, the flock of birds, or the school of fish, since by coherent behavior they can coordinate their movement to find food or to face threats \cite{COUZIN2018844}.   
Also, it is involved in vital functions e.g circadian rhythms \cite{WILSON2022,Cascallares_Gleiser_2015}, the functioning of the heart due to the synchronization of pacemaker cells \cite{Mirollo, YANIV20141210} and,  
physiological brain activity \cite{Fell_Axmacher_2011, Wang-X2010, Buzsaki_2006}. Conversely, abnormal neural synchronization is linked to neurological disorders like epilepsy or Parkinson's disease  \cite{MORMANN2000358, Hammond}.  
\\[2mm]
The archetype to explore the synchronization of a system is the Kuramoto model \cite{Yosiki}, which describes a set of coupled oscillators that interact through a sinusoidal function. Although its simplicity, the Kuramoto model provides a phenomenological description of the problem displaying rich emergent dynamics such as the phase transition from incoherence to synchrony \cite{Gomez, Arenas,RevModPhys_2005}, and provides insight into synchronization process in nature \cite{Odor2,Guo_Zhang_Li_Wang_Yu_2021,Vandermeer_Hajian-Forooshani_Medina_Perfecto,WU2022119002}.  The Kuramoto model has been widely studied and its variations include the presence of noise \cite{Bag,Esfa}, inertia \cite{CHOI201132,Ji_Peron_Rodrigues_Kurths_2014,Florian}, weighted coupling \cite{Li,Wang,Tanaka}, time delay \cite{yeung,PhysRevE.86.016102}, resetting \cite{Tass_1999,Winfree_2001,Gupta_Synch_Reset2022}, among many others \cite{J_Kurths_PhysRep2023,Arenas, RevModPhys_2005,Rodrigues_PhysRep_2016,Dorfler_Bullo_2014}.
\\[2mm]
On the other hand, diverse complex systems suffer a gradual reduction in their capacity to perform specific functions due to the accumulation of damage. Although most of them possess repairing mechanisms, when the damage is severe this reparation process can be incomplete or incorrect due to the urgency of the system to recover without compromising its global performance \cite{wang2009}. A good example of this is the appearance of scars after an injury \cite{Aging_PhysRevE2019}.  The remaining damage or ``misrepairs'' accumulates over time and degrades the functionality of the system \cite{wang2009}.
In the case of complex structures, such as living beings, this phenomenon of damage accumulation can be understood as the mechanism that generates aging. One theory proposed to explain aging in living beings as the accumulation of residual damage resulting from imperfect repair processes, spanning from molecules to tissues. As damage accumulates, it progressively impairs vital functions, making them more vulnerable to environmental hazards, increasing the potential risk of disease,  and eventually culminating in death \cite{wang2009}.  
The whole process of accumulation of damage and degradation of systems can be observed in social systems and civilizations \cite{West}, or in the reduction of the transport capacity in a complex system \cite{Aging_PhysRevE2019,Eraso_Hernandez_2021,Eraso-metro}, just to mention a few examples.
\\[2mm]
The study of dynamical processes in systems with damage accumulation to establish global measures that quantify the degradation of their functionality is important to identify vulnerabilities in systems or to assess the performance of a process. Also, it helps to better understand the relation between the dynamics and the structure of a system to develop new designs robust and resilient to damage. As we mentioned, synchronization process is a universal phenomenon; in particular, its role in vital functions makes it attractive to study in the context of a system with damage accumulation. For example, the synapses between neurons are responsible for the connections in neural circuits, but some dysfunctions can alter their efficiency \cite{MAESTU2015103}. Likewise, intracellular communication may be restricted by aging, potentially influencing the synchronization process of pacemaker cells in the heart \cite{Zeitz_Smyth_2023}. Across various systems, such as power grids, a similar situation arises where transmission lines frequently encounter reduced capacity. This outcome can be attributed to factors like inadequate maintenance, unfavorable weather conditions, and contamination, among others \cite{PANTELI2015259}.  Some researchers have investigated synchronization process in systems that are exposed to severe damage, which implies removing some components of the system. In network science, this effect is modeled with the complete removal of particular groups of links or nodes \cite{WU2022119002, Vasa_Shanahan_Hellyer_Scott_Cabral_Leech_2015,Schumm2020,Tanaka2012,Takeyuki_PlosOne2015,Bonneau_PRE2020,Bonneau_PRE2021}, where the percolation connectivity threshold defines the limit of the global functionality of the system. Furthermore, different authors have explored synchronization in time-varying topologies with modifications in the network structure occurring at a temporal scale comparable to the characteristic time scale of synchronization \cite{GHOSH20221,Zhou_Zou_Guan_Liu_Boccaletti_2016, delGenio_Romance_Criado_Boccaletti_2015, Boccaletti_Hwang_Chavez_Amann_Kurths_Pecora_2006,Porfiri}. A recent work included perturbations in the variables used to characterize the dynamics of oscillators \cite{Wassmer_2021}. As far as we know, synchronization processes in systems that evolve with gradual accumulation of damage have been explored less.
\\[2mm]
In this contribution, we study the influence of cumulative stochastic damage in the synchronization of Kuramoto oscillators on networks.  To this end, we use identical phase Kuramoto oscillators that are coupled through a weighted network where links represent the local coupling of the oscillators, and the accumulation of damage is modeled with a preferential attachment mechanism where edges with previous faults are more likely to be damaged. Also, the damage affects the structure in a non-symmetric way and decreases the coupling strength of the oscillators, provoking alterations in the capacity of the system to reach synchronization from random phases. We analyze different structures that fully synchronize in the absence of any damage and have different topologies. We include two deterministic networks as a wheel and the lattice formed by the movements of the knight on a chess board that contrast with two random networks generated with the  Erd\H{o}s-R\'enyi and Barab\'asi-Albert algorithms. Additionally, we analyze 109 non-isomorphic networks with six nodes. Our findings allow us to classify the effect of damage in networks. The explored framework is general and introduces different tools for the characterization of the effect of accumulation of damage in diverse dynamical processes in complex systems.   
\section{Kuramoto model  of identical oscillators}
\subsection{Preliminaries}\label{general}
\begin{figure*}[t]
	\centering
	\includegraphics*[width=1.0\textwidth]{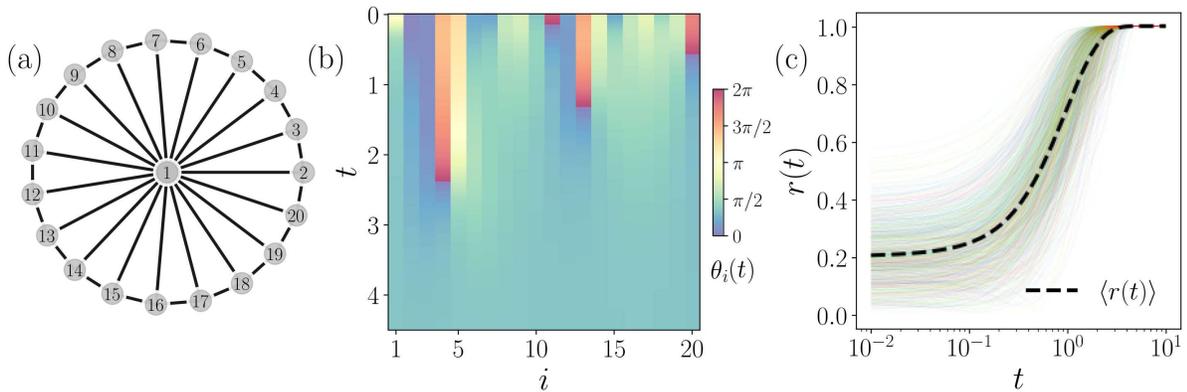}
	\vspace{-6mm}
	\caption{\label{Fig_1} Synchronization processes in a wheel graph with $N=20$ nodes.  (a) Wheel graph. (b) Temporal evolution of the phases $\theta_i(t)$ as a result of the numerical integration of Eq. (\ref{inikuramoto}) using random initial phases sampled from a uniform distribution in the interval $[0,2\pi)$, the value of the phases is codified in the color bar. (c) Evolution of the order parameter $r(t)$ in the synchronization of the network. Thin lines represent the results of $1000$ realizations using random initial phases and the dashed line shows the ensemble average denoted as $\langle r(t)\rangle$.}
\end{figure*}

We consider a system with $N$ nodes formed by identical  Kuramoto oscillators with a coupling structure defined by a network where the natural frequency $\omega$ is the same for each oscillator \cite{Taylor_2012}. However, all the methods developed can be adapted to the analysis of systems where the oscillators have different characteristics; for example, in systems where natural frequencies are sampled from a probability distribution. At a time $t$, the oscillators are characterized by their phases $\theta_i(t)$ with $i=1,2,\ldots,N$ \cite{Arenas,RevModPhys_2005,Rodrigues_PhysRep_2016}. The oscillators in the original Kuramoto model are all-to-all coupled  \cite{Yosiki}; however, the coupling can be defined using the connectivity information of a network \cite{Arenas,RevModPhys_2005}. This coupling network corresponds to a  simple connected graph formed by a set of $N$ nodes denoted by $\mathcal{V}$ and a set of edges $\mathcal{E}$  that connect pairs of nodes $(i,j)$,  $|\mathcal{E}|$ is the total number of different edges in the network. The topology of the structure is described by its $N\times N$ adjacency matrix $\mathbf{A}$  with elements $A_{ij}=A_{ji}=1$ if the nodes $i,j$ are connected and $A_{ij}=0$ otherwise; the diagonal elements are $A_{ii}=0$ since the nodes are not connected with themselves. In this system,  the evolution of phases of identical Kuramoto oscillators placed in the nodes is described by the system of nonlinear coupled differential equations  \cite{Taylor_2012}

\begin{equation}\label{normalkuramoto}
     \frac{d\theta_i(t)}{dt}=\omega+K\sum_{j=1}^{N}A_{ij}\sin[\theta_j(t)-\theta_i(t)],
\end{equation}
for $i=1,2,\ldots,N$ and, where $K>0$ corresponds to the global coupling strength of the system. In addition, the system in Eq. (\ref{normalkuramoto}) satisfies rotational symmetry so it is invariant under the transformation $\theta_i\to\theta_i+\omega t$. In this way, it is possible to set $\omega=0$ and rescale the time by setting $K=1$ \cite{Townsend}. Then, Eq. (\ref{normalkuramoto})  is transformed into \cite{Taylor_2012,Townsend,Jadbabaie_Motee_Barahona_2004}
\begin{equation}\label{inikuramoto}
    \frac{d\theta_i(t)}{dt}=\sum_{j=1}^{N
    }A_{ij}\sin[\theta_j(t)-\theta_i(t)].
\end{equation}
%
The Kuramoto system in Eq. (\ref{inikuramoto}) is a gradient system that eventually evolves to a fixed point \cite{Taylor_2012, Townsend,Jadbabaie_Motee_Barahona_2004}. Some of the equilibrium states correspond to coherent states, which means all the oscillators in the system take the same phase value. In other words, the system is completely phase synchronized \cite{Taylor_2012}. The conditions under which a Kuramoto system evolves to complete synchronization are still under study but the topology of the system plays an important role \cite{Taylor_2012,Townsend,Ling, HA20101692}.  A way to assess the phase coherence of the Kuramoto oscillators  is  through the macroscopic order parameter \cite{Arenas, RevModPhys_2005,Rodrigues_PhysRep_2016}
\begin{equation}\label{orderparam}
    r(t)=\frac{1}{N}\left|\sum_{j=1}^{N}  \exp\left[\mathbf{i}\theta_j(t)\right]\right|,
\end{equation}
where $\mathbf{i}=\sqrt{-1}$. From Eq. (\ref{orderparam}), $0\leq r(t)\leq 1$. In particular, if the Kuramoto system  exhibits	complete phase coherence $r(t)= 1$, on the contrary  $r(t)= 0$ if the oscillators move in a completely incoherent manner or form  clusters of synchronized oscillators. 
\\[2mm]
As an example of a Kuramoto system that evolves to a completely synchronized state,  we present in Fig. \ref{Fig_1} the evolution of $N=20$ identical Kuramoto oscillators arranged on a wheel graph.  A $N$-vertex wheel graph is formed by  $N-1$ nodes arranged on a ring and connected to a common vertex. Fig. \ref{Fig_1}(a) shows the wheel graph representing the connectivity network of the system.  The result in Fig. \ref{Fig_1}(b) displays the evolution of phases $\theta_i(t)$ for each oscillator $i$, ($i=1,2,\ldots,20$) as a function of time $t$, the values of the phases are encoded in the color bar. In this case, the system starts with random initial phases distributed uniformly on the interval $[0,2\pi)$ and eventually all of them reach the same value. In Fig. \ref{Fig_1}(c) we present the behavior of the order parameter $r(t)$, the different curves represent $1000$ realizations of the process and the dashed line describes the ensemble average $\langle r(t)\rangle$.  Initially, $r(t)$ takes different values between 0 and 0.6 due to the random distribution of the phases at $t=0$, then $r(t)$ evolves for all the conditions to  1 as the oscillators reach a coherent state. This general behavior is also observed in the ensemble average $\langle r(t)\rangle$.
\\[2mm]
Despite the simplicity of the modeling of synchronization with identical oscillators, Eq. (\ref{inikuramoto}) includes the topology of the network, so this framework is appropriated to deal with the problem of the influence of topology in the dynamics of coupled oscillators \cite{Taylor_2012,Jadbabaie_Motee_Barahona_2004}. Additionally, we consider that the presence of stochastic terms or different frequencies of heterogeneous oscillators can potentially mask the effects of the phenomenon we want to study.
In the following, we are interested in the global changes in the synchronization process due to the effects of damage in the links of the coupling network; therefore, this model of identical Kuramoto oscillators results in a convenient starting point.
\subsection{Synchronization times}
\label{time}
The evolution to a completely synchronized state in systems of identical Kuramoto oscillators means that all the oscillators acquire identical phases. Strictly, the time at which these systems achieve a fully synchronized state is infinite; nevertheless, it is possible to relax that condition \cite{Almendral_2007}. For the sake of convenience, we analyze the time at which the systems are almost synchronized and the order parameter $r(t)$ takes a determined fixed value $r$ close to 1, we denote this synchronization time as $\tau_0$. From Eq. (\ref{inikuramoto}), we observe that $\tau_0$ is a variable that depends on the topology of the network and the initial phases of the system, if we consider random phases at $t=0$,   $\tau_0$ can be seen as an stochastic variable and the statistical analysis of this quantity gives us relevant information of the system. 
\\[2mm]
\begin{figure*}[t!]
	\centering
	\includegraphics*[width=1.0\textwidth]{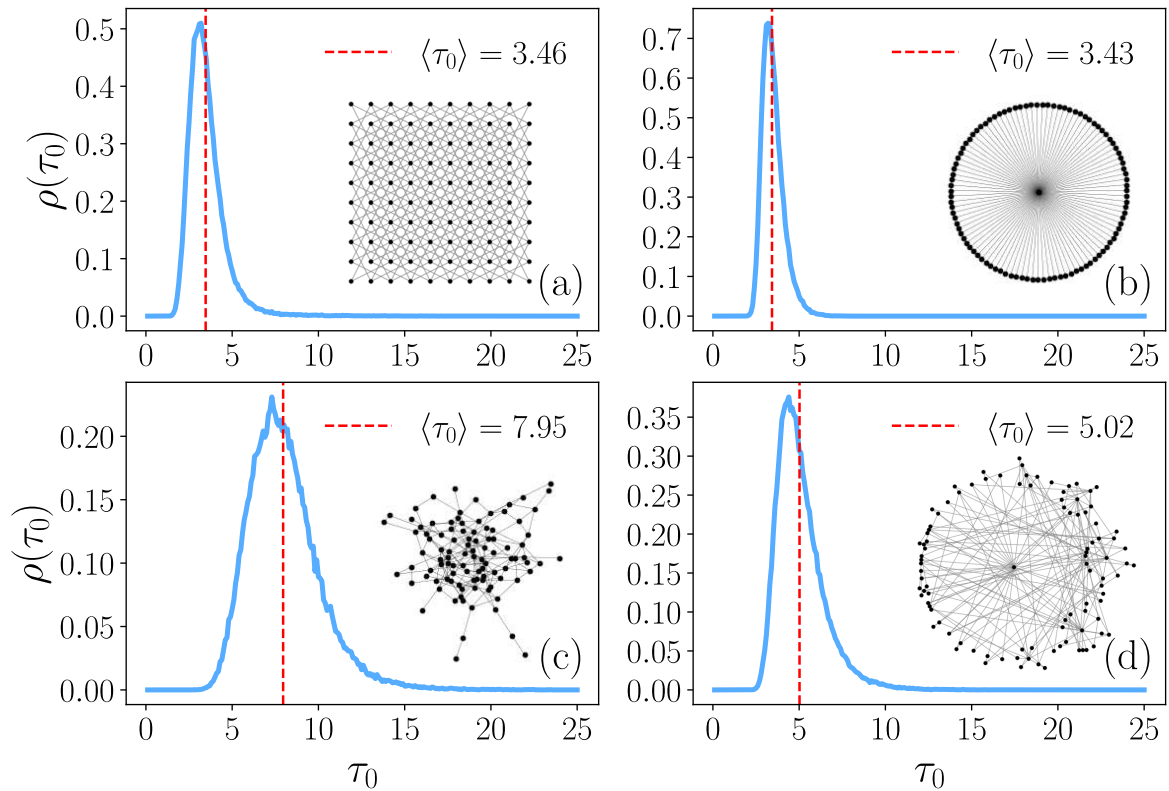}
	\vspace{-6mm}
	\caption{\label{Fig_2}Probability densities $\rho(\tau_0)$ of synchronization times $\tau_0$ for networks with $N=100$ nodes. (a) A knight network with dimensions $10\times 10$,  (b) a wheel, (c)  an Erd\H{o}s-R\'enyi network with $p=0.04$, (d) a Barab\'asi-Albert network with $m=2$. The results are generated using $10^5$ Monte Carlo simulations of the model in Eq. (\ref{inikuramoto}) considering random initial phases drawn from a uniform distribution in the interval $[0,2\pi)$ and evaluating through numerical integration the time $\tau_0$ when the threshold  $r=0.99$ is crossed from below. Vertical dashed lines represent the ensemble average $\langle \tau_0\rangle$ and each network analyzed is presented as an inset.  We use bins with size $\Delta \tau_0=0.1$ in the calculation of the probability densities. }
\end{figure*}
In the following, we define $\tau_0$ as the time at which the system has reached a state such that the threshold $r=0.99$ is crossed from below. To find these times, we perform numerical integration of Eq. (\ref{inikuramoto}) with random initial phases sampled from a uniform distribution in the interval $[0,2\pi)$. We use the fourth order Runge-Kutta method \cite{DORMAND198019} implemented in C++ in the GNU Scientific Library \cite{galassi2018scientific}. Here, it is important to highlight that several works have shown that in the numerical solution of Eq. (\ref{inikuramoto}), the results do not depend significantly on the integration method implemented or on the time step $\Delta t$ used to obtain numerical solutions \cite{Odor2,Muller,Bohle}.
\\[2mm]
In Fig. \ref{Fig_2}, we present the probability density of synchronization times $\tau_0$ for different networks with $N=100$ nodes. The probability densities were generated from $10^5$ realizations of $\tau_0$ obtained using random initial phases. In Fig. \ref{Fig_2}(a), we analyze the synchronization times for a  particular lattice called knight graph formed by all possible movements of a knight chess piece in a chessboard with dimensions $10\times 10$ \cite{Watkins_2012}. In Fig. \ref{Fig_2}(b), we show the results for a wheel graph. Panels  \ref{Fig_2}(c)-(d) show the analysis of two random networks with completely different topologies.  Figure \ref{Fig_2}(c) depicts our findings for an  Erd\H{o}s-R\'enyi network constructed starting from a set of $N=100$ of nodes which are joined by edges whose ends are selected at random with probability $p=0.04$ among all the vertices \cite{Erdos:1959:pmd}, in this model the degree distribution can be approximated by the Poisson distribution \cite{VespiBook}. Finally, Fig. \ref{Fig_2}(d) presents the results for a Barab\'asi-Albert network generated using a  preferential attachment algorithm so each new node is connected to $m=2$ nodes chosen accordingly to their number of connections or degree  \cite{Barabasi}, this algorithm provides an example of the emergence of networks with heavy-tailed degree distributions in terms of the elementary process governing the wiring of new vertices joining the network \cite{VespiBook}. All the networks explored are plotted as insets and vertical dashed lines show the ensemble average for the values $\tau_0$.
\\[2mm]
The probability densities $\rho(\tau_0)$ show that the $\tau_0$ are close to certain values but the curves are asymmetric. The different curves evidence the influence of the topology of the network in the synchronization process. For each network, we notice that characteristics such as the width and height of the curves are different. It is known that there is an effect of the structure of the network to achieve synchronization \cite{Lacerda_Freitas_Macau_Kurths_2021,Barahona_Pecora_2002, Gomez_Gardenes} and the results show that this effect can be observed in the probability densities of $\tau_0$. Concerning the mean value  $\langle \tau_0\rangle$  for each network, we notice that this quantity is different for all of them, Fig. \ref{Fig_2}(a) and Fig. \ref{Fig_2}(b) exhibit results that differ slightly; however, in Figs. \ref{Fig_2}(c) or (d) the differences are more obvious as a consequence of the relation of $\langle \tau_0\rangle$    with the connectivity of the networks \cite{Almendral_2007}. These results show that $\tau_0$ is a good candidate to study the performance of synchronization of a network of identical Kuramoto oscillators. In the following part, we will see that this quantity becomes relevant when we explore the synchronization of networks with damage.
\section{Synchronization of systems with accumulation of damage}
\subsection{Modeling the damage on complex systems}\label{damage}
In this section, we present a brief overview of the method developed in Refs.  \cite{Aging_PhysRevE2019,Eraso_Hernandez_2021} to model the process of generation of cumulative damage and aging in networks. In general, the process of accumulation of damage reduces the functionality of the links of the network with fails generated stochastically  with a preferential attachment mechanism. The characteristic time scale of this process of accumulation of damage is represented by time $T$.
The time interval at which system receives damage $\Delta T$ is long compared to the characteristic time of other dynamical processes that can take place in the system such as a transportation or synchronization process, for $\Delta T=1$ (measured in units of the system damage accumulation), $T=0,1,2,\ldots$. We see for example this difference in the time scales in the activity of a metro system, where the daily activity of users in trains is measured in hours or days but the damage of the infrastructure is only observed at the scale of months or years. We refer the reader to Ref. \cite{kuehn2015multiple} for a formal mathematical treatment of systems that exhibit multiple time scale dynamics.
\\[2mm]
The damage in the edge $(i,j)$ of the network is generated at time $T$ with probability  $\pi_{ij}(T)$ \cite{Aging_PhysRevE2019}
\begin{equation}\label{problinks}
\pi_{ij}(T)=\frac{h_{ij}(T-1)}{\sum_{(l,m) \in \mathcal{E}} h_{lm}(T-1)}\qquad (i,j) \in \mathcal{E},
\end{equation} 
where $h_{ij}(T)$ is a stochastic integer variable such as $h_{ij}(T)-1$ counts the number of
random faults that exist in  link $(i,j)$ at time $T$.  Initially,  at $T=0$, $h_{ij}(0)=1$, i.e. there are no faults in any edge in the network. At this point, the algorithm uniformly selects an edge. After damage occurs in any edge, $T$ is increased by 1 and the probabilities of the edges to receive a new fault change according to Eq. (\ref{problinks}). The entire process generates structures with preferential distribution of damage, where  new faults are more likely to accumulate in the most affected edges. Since the damage in edge $(i,j)$ is independent of the damage that receives $(j,i)$, the method generates  an  asymmetric  $N\times N$ matrix of weights $\mathbf{\Omega}(T)$ describing the couplings between oscillators. $\mathbf{\Omega}(T)$ is introduced to characterize the state of damage of the connections of the system and its elements  are defined by \cite{Aging_PhysRevE2019,Eraso_Hernandez_2021}
\begin{equation}\label{OmegaijT}
\Omega_{ij}(T)=(h_{ij}(T))^{-\alpha} A_{ij}
\end{equation}
where $\alpha\geq 0$ is a real value called the ``misrepair" parameter that quantifies the response of the system to damage in terms of repair capacity \cite{Aging_PhysRevE2019}; in this manner, $\alpha$ shows how detrimental the accumulation of damage is. On the one hand, in the limit $\alpha\to 0$, the system is able to repair completely after receiving damage, in such a way that  $\Omega_{ij}(T) \to A_{ij}$. In the limit $\alpha\to\infty$,  the system can not be repaired so a hit in a link is equivalent to its removal from the network, this removal of edges gradually brings the system to the percolation limit, where the network becomes disconnected.
 \\[2mm]
This aging process has been used to analyze the transport capacity of networks under stochastic damage and has allowed researchers to conclude that more complex structures are more resilient to the process of aging \cite{Aging_PhysRevE2019,Eraso_Hernandez_2021}. A recent application is the use of this model to study the robustness of metro systems under damage accumulation \cite{Eraso-metro}. In the following part, we study the effects of damage accumulation on networks whose main function is to synchronize. In particular, we are interested in cases where the system is able to repair but not in a perfect way so it starts to accumulate damage in its links, which means that $\alpha$ only takes finite values.
\subsection{Functionality}\label{functionality}
In Sec. \ref{general}, we introduced the  Kuramoto model of identical oscillators in a network without damage. Let us now include the damage in the network and introduce a measure to quantify its consequences in synchronization.   We consider that damage modifies the connectivity of the underlying structure, generating a  reduction of coupling between the oscillators in the network. In this new structure, the Kuramoto system receives damage in a time interval $\Delta T$ which is considerably large in comparison to the time scale $t$ of the synchronization process of the network, $\Delta T$ is enough time to the system to synchronize from an incoherent state. The variable $T$ represents the time at the scale of damage generation. But due to the fact that  $T$ increases by 1 once the system receives new damage, $T$ also provides information on the amount of global damage that the system has received. The matrix of weights  $\mathbf{\Omega}(T)$, with elements defined in Eq. (\ref{OmegaijT}), plays the role of a weighted connectivity matrix in the initial Kuramoto model. In this manner,  for each configuration of damage characterized by $T$, the system of nonlinear equations that define the dynamics of the Kuramoto oscillators is given by
\begin{equation}\label{Kuramoto}
   \frac{d\theta_i(t)}{dt}=\sum_{j=1}^{N
    }\Omega_{ij}(T)\sin[\theta_j(t)-\theta_i(t)].
\end{equation}
Linear models are commonly used to study synchronization. A linear approximation of the Kuramoto model is valid if the system operates close to synchronization. In such scenarios, the evolution of the system can be effectively studied in terms of the eigenvalues and eigenvectors of the Laplacian matrix of the network. To complement this section,  the linear approximation is discussed in detail in Appendix \ref{Append1}. However, we are interested in the general dynamics described by Eq. (\ref{Kuramoto}).
\\[2mm]
Several works have studied the influence of time-varying couplings in the synchronization process of networks in scenarios where the changes of the coupling network can be fast or slow in comparison to the synchronization times of the system \cite{GHOSH20221,Zhou_Zou_Guan_Liu_Boccaletti_2016, delGenio_Romance_Criado_Boccaletti_2015, Boccaletti_Hwang_Chavez_Amann_Kurths_Pecora_2006}. In that framework, the effects of the network alteration are evaluated by means of stability analysis of the synchronization state. We implement a different approach assuming the dynamics as a multiple time scale process for which the changes produced by the damage occur at a larger scale in comparison with the synchronization times.
\\[2mm]
In the following, we assume that the function of the system at each configuration of damage described by  $\mathbf{\Omega}(T)$ is to reach the global synchronization starting from random phases. In order to evaluate this capacity of synchronization on networks with damage at each $T$, we propose a measure of ``functionality'' $\mathcal{F}_{\mathrm{s}}(T)$ that quantifies the ``health'' of the system comparing the synchronization times in Sec. \ref{time}, at different stages of damage. We define
\begin{equation} \label{Fun}
    \mathcal{F}_{\mathrm{s}}(T)\equiv\frac{\tau_0}{\tau(T)}.
\end{equation}
Here, we denote as $\tau(T)$ to the synchronization time when the network has suffered damage with couplings described by $\mathbf{\Omega}(T)$, the initial phases at $t=0$ are chosen randomly but remain the same when evaluating $\tau_0$ and $\tau(T)$.  The functionality $\mathcal{F}_{\mathrm{s}}(T)$ in Eq. (\ref{Fun}) is a global measure of the effects of damage in the synchronization process on networks. Particularly  $\mathcal{F}_{\mathrm{s}}(0)=1$ and on average decreases with $T$. In situations when the network fails to synchronize after a certain amount of damage has been added $\tau (T) \to \infty$. Consequently, the network is incapable of carrying out its function leading to $\mathcal{F}_{\mathrm{s}}(T)\to 0$.       
\subsection{Effects of  damage in synchronization process}
\begin{figure}[t]
	\centering
	\includegraphics*[width=1\textwidth]{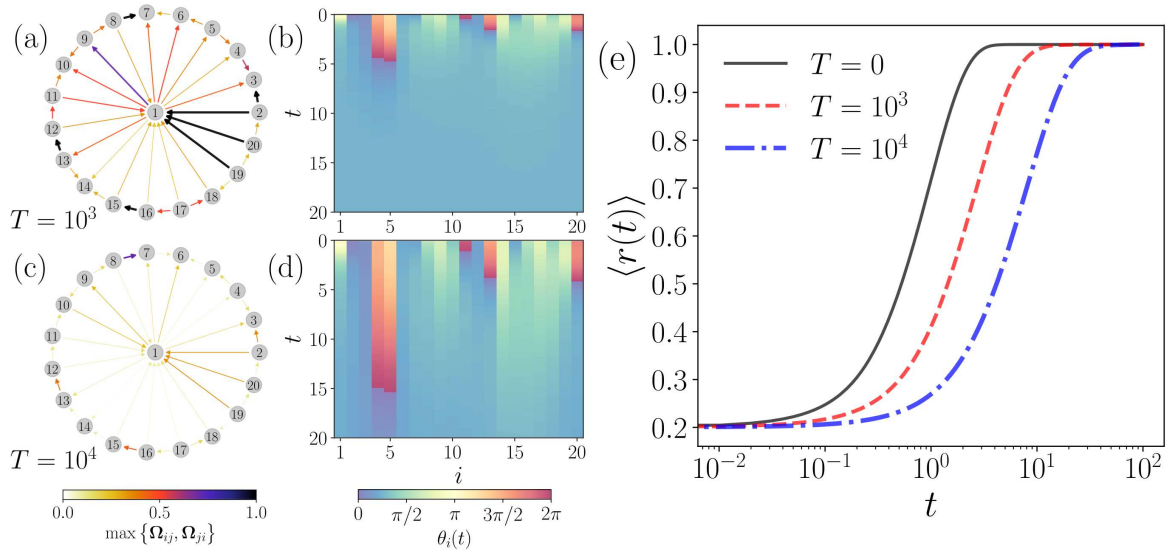}
	\vspace{-5mm}
	\caption{\label{Fig_3} Synchronization process in a wheel with $N=20$ nodes with damage at  $T=10^3$ and $T=10^4$ with $\alpha=0.5$. Panel (a) shows the network and its weighted connections at $T=10^3$, only the links with the largest weight are presented, values are codified in the color bar. Panel (b) depicts the evolution of the phases as a function of $t$ at $T=10^3$.  The color represents the value of the phases $\theta_i(t)$ codified in the color bar. Panels (c) and (d) present the state of the network at $T=10^4$ and its evolution. Panel (e) displays the evolution of the ensemble average  of the order parameter $\langle r(t)\rangle$  for the network at the different scenarios of damage,  average values are obtained considering $1000$ realizations.}
\end{figure}

In this section, we study the effects of damage on the global synchronization of Kuramoto systems. To illustrate this, in Fig. \ref{Fig_3} we present the behavior of synchronization of a wheel graph with $N=20$ nodes for different values of $T$ characterizing the damage and $\alpha=0.5$. In Fig. \ref{Fig_3}(a) we depict the network with its weighted links at $T=10^3$. The network is drawn showing only the heaviest edge between each pair of connected nodes, and the weights of the links are codified according to the color bar. We can observe how some of the links keep their original weight while others are notably affected as a consequence of the preferential damage distribution. In Fig. \ref{Fig_3}(b) we present the evolution of the phases of each oscillator at $T=10^3$.  The system starts with the same initial conditions that we used in the example in Fig. \ref{Fig_1}(b), after some steps the phases reach a coherent state; however, we can observe that in this case, the time to achieve the coherent state is longer than the time used in the network without damage presented in Fig. \ref{Fig_1}(b). Fig. \ref{Fig_3}(c) shows the network at $T=10^4$, in this case, we observe a very deteriorated network with connections that almost disappear. Fig. \ref{Fig_3} (d) depicts the evolution of phases of the system at $T=10^4$, in a similar way to the case with $T=10^3$, the system starts with the same initial conditions and evolves to a synchronized state, but now the time needed to achieve this configuration is still longer than in the cases with less damage. 
\\[2mm]
On the other hand,  Fig. \ref{Fig_3}(e) displays the ensemble average of the order parameter $\langle r(t)\rangle$ for $T=0$, $T=10^3$ and $T=10^4$. In each case, the average is over $1000$ realizations of Monte Carlo simulations, and the set of random initial conditions used to evaluate $r(t)$ in each realization at $T=0$ is the same for the cases $T=10^3$ and $T=10^4$. The curves show the general behavior of the system under different random initial conditions and diverse stages of damage. The results show that on average, the system evolves from an incoherent initial state to a synchronized state. Initially, all the curves remain close to each other and $\langle r(t)\rangle$ does not change significantly, eventually, the curves separate and evolve at different rates so each one reaches 1 at a different time. The curve for $T=0$  arrives first, then the curve for $T=10^3$ and finally the curve for $T=10^4$. This example shows a system with complete synchronization and how the damage in the coupling network modifies the time in which the system reaches that state of full coherence.
\subsection{Functionality reduction in the Kuramoto model}
 \begin{figure}[t]
	\begin{center}
		\includegraphics*[width=1\textwidth]{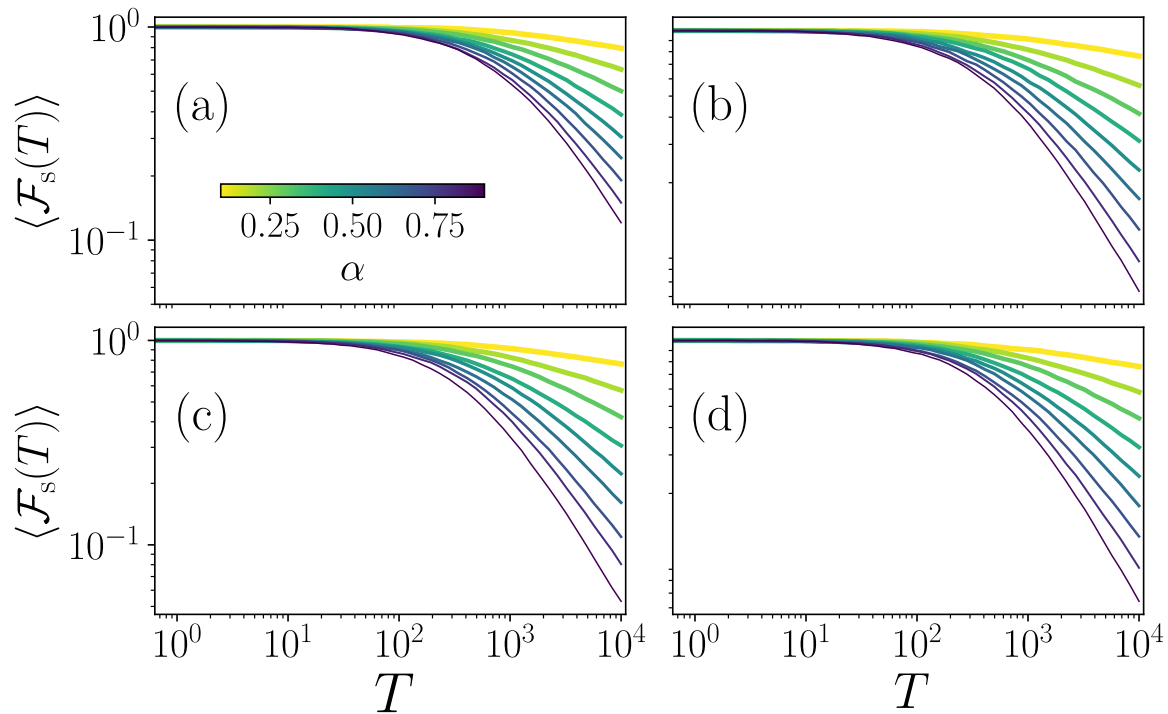}
	\end{center}
	\vspace{-5mm}
	\caption{\label{Fig_4} Reduction of the ensemble average of the functionality of synchronization $\langle\mathcal{F}_{\mathrm{s}}(T)\rangle$ as the damage increases in networks of different topologies with $N=100$.
		The curves correspond to $\langle\mathcal{F}_{\mathrm{s}}(T)\rangle$ for $1000$ realizations and different values of $\alpha$ encoded in the color bar. We analyze the networks in Fig. \ref{Fig_2}: (a) Knight graph of dimension $10\times10$, (b) a wheel, (c)  an Erd\H{o}s-R\'enyi network and, (d) a Barab\'asi-Albert network. The Monte Carlo simulations were performed for $1000$ realization using random initial phases sampled from  a uniform distribution in the interval $[0,2\pi)$.}   
\end{figure}
In Sec. \ref{functionality}, we introduced a measure of functionality of the synchronization  $\mathcal{F}_{\mathrm{s}}(T)$ which gives us insight into the response of the synchronization process under the damage of the network and it can be used to compare networks with different structures.   In Fig. \ref{Fig_4} we present the evolution of the ensemble average of the functionality $\mathcal{F}_\mathrm{s}(T)$  defined in Eq. (\ref{Fun}) as a function of $T$ for the networks explored in Fig. \ref{Fig_2} and different values of $\alpha$. For the evaluation of  $\mathcal{F}_{\mathrm{s}}(T)$ we use the numerical values of $\tau_0$ and $\tau(T)$ obtained when the system reaches a  coherent state with a fixed order parameter $r=0.99$ (see the appendix in Sec. \ref{Append2} for a detailed discussion of the effect of having different threshold values $r$). The averages were calculated over $1000$ realizations.  The different values of $\alpha$  are codified in the color bar that appears in Fig.  \ref{Fig_4}(a). In Fig. \ref{Fig_4}(a) we present the results for the knight graph, in Fig. \ref{Fig_4}(b) the wheel graph, in Fig. \ref{Fig_4}(c) the Erd\H{o}s-R\'enyi network and in Fig. \ref{Fig_4}(d) the results for the  Barab\'asi-Albert network. In all the cases we observe that $\langle\mathcal{F}_{\mathrm{s}}(T)\rangle$ is closer to 1  when the damage in the system is small nevertheless, as the damage increases, the functionality decreases.   These results are in agreement with previous findings in Fig. \ref{Fig_3}, showing the increase of the synchronization time $\tau(T)$. Also, the changes in $\alpha$ affect the functionality since the increase in $\alpha$ implies an increase in the intensity of damage so for large values of $\alpha$ the loss of functionality is more significant.  One of the interesting features that we find in this result is the similarity observed in Figs. \ref{Fig_4}(b)-(d). Despite the topological differences of the networks, including the different degree distributions,  the results of the average ensemble are identical, and the case depicted in  Fig. \ref{Fig_4}(a) is just a re-scaled version. The reason behind this result is the number of edges of the analyzed network. As the damage is distributed across the edges, part of the robustness exhibited by these structures against damage comes from the number of edges. The knight graph has $576$ edges, the wheel has $396$, and the Erd\H{o}s-R\'enyi and Barab\'asi-Albert networks have $392$ edges. The ensemble average of the functionality of synchronization scales with the number of edges of the graphs. In the next section, we continue to study the properties of this measure, especially its relationship with the topology of the networks. 

\section{Synchronization on small graphs with damage}
\begin{figure}[t]
	\begin{center}
		\includegraphics*[width=1.0\textwidth]{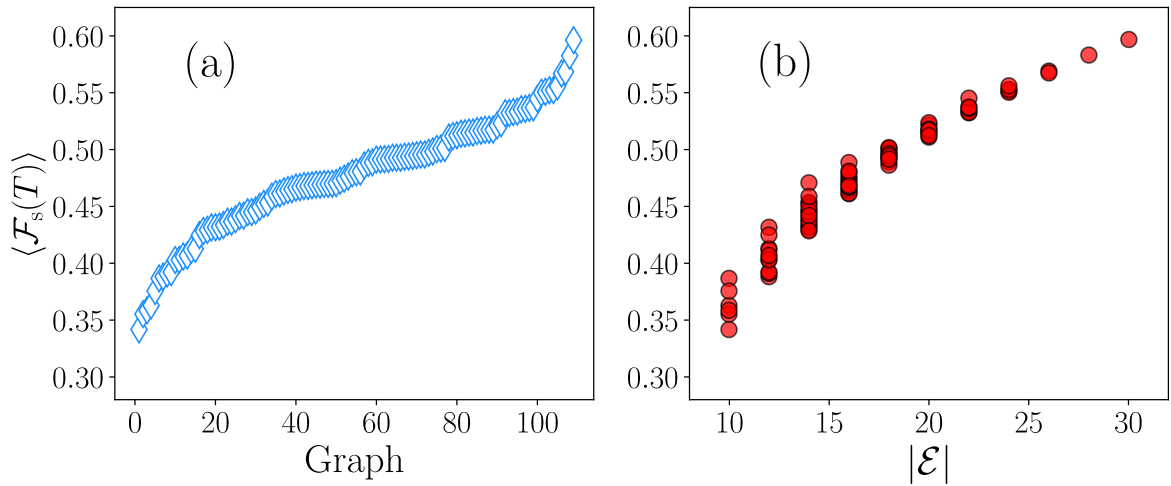}
	\end{center}
	\vspace{-6mm}
	\caption{\label{Fig_5} Ensemble average functionality $\langle\mathcal{F}_{\mathrm{s}}(T)\rangle$ for cumulative damage with $\alpha=0.5$ and $T=100$ for  109 non-isomorphic connected graphs with $N = 6$ nodes. (a) Graphs sorted with the values of $\langle\mathcal{F}_{\mathrm{s}}(T)\rangle$. (b) $\langle\mathcal{F}_{\mathrm{s}}(T)\rangle$
		as a function of the number of directed edges $|\mathcal{E}|$. The values are obtained considering $10^7$ realizations. See details in the main text.
	}
\end{figure}

Having defined synchronization on networks under the influence of accumulation of damage, as well as introduced the concept of functionality $\mathcal{F}_{\mathrm{s}}(T)$ in Eq. (\ref{Fun}), and its reduction with $T$; in this section, we explore the relation between $\langle\mathcal{F}_{\mathrm{s}}(T)\rangle$ and the structure of the network. To this end, we study the effect of damage on all the connected non-isomorphic graphs with size $N=6$, the set of graphs analyzed is available in Ref. \cite{ConnectedGraphs} and contains 112 graphs providing a great variety of topologies that includes several trees (for example, the linear graph or the star graph), networks with cycles with different lengths and structures with a high density of edges including the fully connected graph. Also, having only six nodes, we can generate more realizations of the Monte Carlo simulations to perform a better statistical analysis of the results.
\\[2mm]
In the following, we are interested only in structures that fully synchronize in the absence of damage, thus we omit three particular graphs (a ring with six nodes and two networks formed by a ring with five nodes added one and two edges). In this manner, we consider 109 graphs in the dataset. For each network, we calculate the synchronization times $\tau_0$ and $\tau(T)$, defined as the minimum time $t$ for which $r(t)$ in Eq. (\ref{orderparam}) reach the value $r=0.99$ in the synchronization on the original network (to obtain $\tau_0$) and the structure with accumulation of damage with $T=100$ and $\alpha=0.5$ (to obtain $\tau(T)$). We generate $10^7$ pairs ($\tau_0$, $\tau(T)$) using initial random phases. However, we maintain the same initial condition to calculate $\tau_0$ and $\tau(T)$ in each realization. The results are obtained using numerical integration of Eqs. (\ref{Kuramoto}).
\\[2mm]
Once we generated the set of values ($\tau_0$, $\tau(T)$) for each graph, we can evaluate the ensemble average of the functionality $\langle\mathcal{F}_{\mathrm{s}}(T)\rangle=\langle \tau_0/\tau(T)\rangle$. Our findings are presented in Fig. \ref{Fig_5} where, in panel \ref{Fig_5}(a), the graphs are sorted in increasing values of $\langle\mathcal{F}_{\mathrm{s}}(T)\rangle$ ranging from the most affected by damage with $\langle\mathcal{F}_{\mathrm{s}}(T)\rangle=0.34175$ (associated to the star graph) to the structure that better tolerates the damage with $\langle\mathcal{F}_{\mathrm{s}}(T)\rangle=0.59621$ (corresponding to the fully connected graph). The results show variations in the values of the functionality in the set of graphs analyzed, in some cases with differences $\langle\mathcal{F}_{\mathrm{s}}(T)\rangle$ between two graphs at the order of $10^{-5}$; however, such variations can be identified with $10^7$ realizations. In addition, in panel \ref{Fig_5}(b) we plot $\langle\mathcal{F}_{\mathrm{s}}(T)\rangle$ as a function of the total number of edges $|\mathcal{E}|\equiv\sum_{i=1}^N\sum_{j=1}^N A_{ij}$ (considering both directions in each edge). In this visualization of the results, we see that the total number of edges is an important quantity to determine if the network is resilient to cumulative damage; in particular, the higher $|\mathcal{E}|$, the damage is distributed in more edges making them be less affected in its functionality. However, the relation between $\langle\mathcal{F}_{\mathrm{s}}(T)\rangle$ and $|\mathcal{E}|$ is non-linear and, for a fixed value $|\mathcal{E}|$, subtle differences appear associated to the particularities in the topology of each network.
\\[2mm]
\begin{figure*}[ht!]
	\begin{center}
		\includegraphics*[width=1\textwidth]{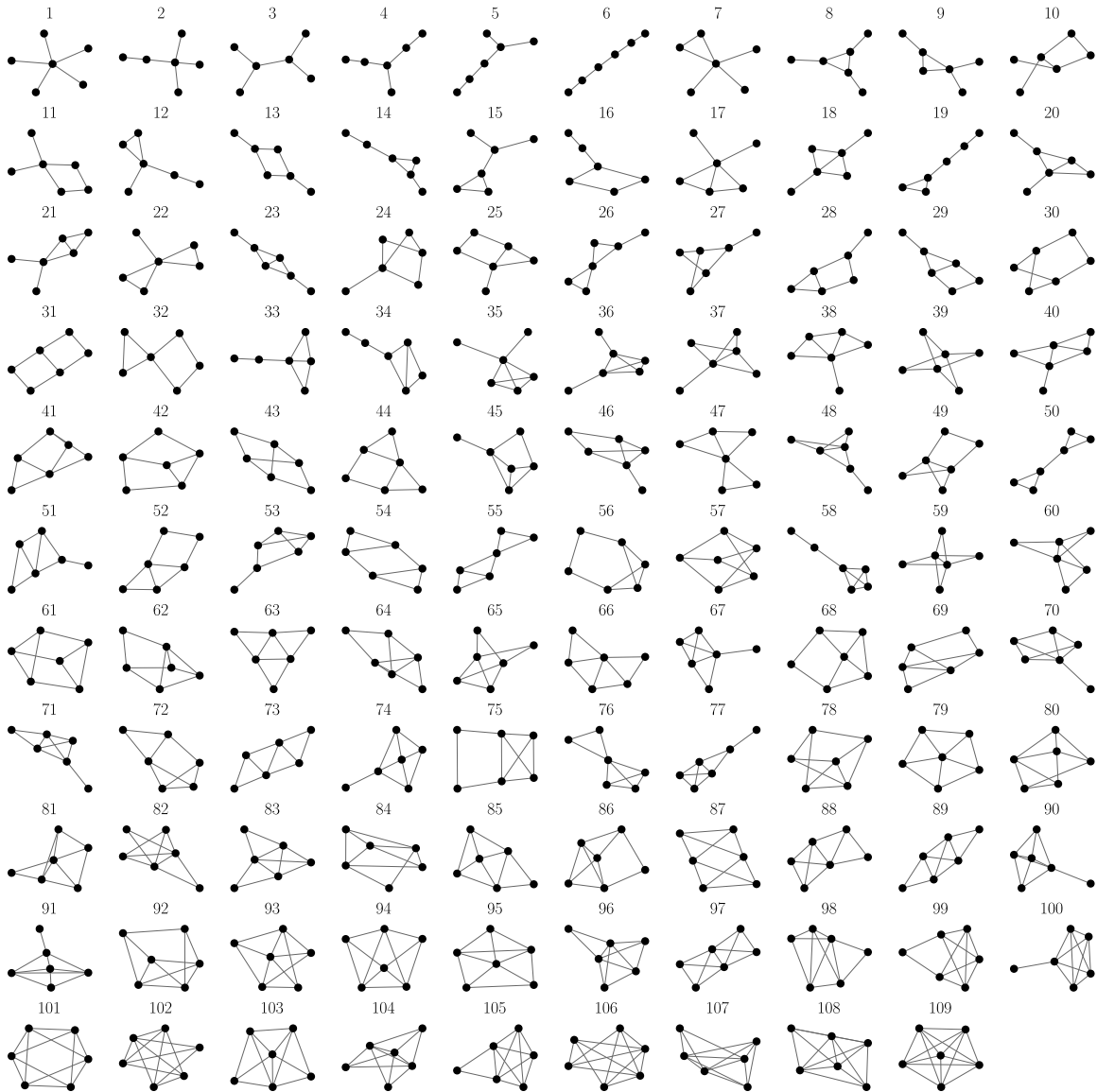}
	\end{center}
	\vspace{-5mm}
	\caption{\label{Fig_6} Non-isomorphic connected graphs with $N=6$ that synchronize at finite times. The networks are sorted according to their response to damage considering the value of $\langle\mathcal{F}_{\mathrm{s}}(T)\rangle$ at $T=100$ presented in Fig. \ref{Fig_5} for $\alpha=0.5$ and $10^7$ realizations of synchronization on damaged networks. The graphs are obtained from \cite{ConnectedGraphs}. Of the  112 graphs in the original dataset, 3 graphs were discarded because they do not reach complete synchronization for several initial conditions. }
\end{figure*}
In Fig. \ref{Fig_6}, we present all the 109 graphs  sorted according to the values of $\langle\mathcal{F}_{\mathrm{s}}(T)\rangle$ at $T=100$, as in Fig. \ref{Fig_5}. This classification starts with tree networks (1 to 6) with the lowest functionality values showing reduced tolerance to damage in the synchronization process. The following networks are more connected and have structures such as cycles and more redundant paths. 
Something important to highlight in the results is that this order obtained for synchronization differs from a classification obtained in Ref. \cite{Aging_PhysRevE2019} for networks with damage but with a global functionality associated with the capacity of random walkers to reach any node of the network. For example, in graph 50 called barbell (formed by two triangles and an edge that joins them), in the transport process, this structure is more fragile since the removal or damage of the edge that joins the two triangles alters drastically the connectivity of two important parts of the network affecting the global transport. In contrast, the implementation of the functionality in Eq. (\ref{Fun}) considers the couplings between neighbors and global synchronization. Therefore, in comparison with the transport, the synchronization in the barbell graph is not severely affected by the fragility of a link connecting the two triangles, since each group of nodes is independently resistant to damage.
This shows that in the effects of accumulated damage the mechanisms that make a network more resilient in transport issues are not the same when the function of the network is to reach states with complete synchronization.
\\[2mm]
On the other hand, one aspect that stands out among the structures with the same number of edges is the fact that the most resistant topologies have a  subgraph with higher connectivity and linear chains of one or two nodes with one of their ends fixed to the most connected structure. This case is observed, for example, in networks 19 and 34. One possible explanation is that such a mechanism receives little damage in the parts with fewer connections and it is more likely to concentrate damage in the subgraph with higher connectivity. This group of nodes operates as a regulator that prevents from damage key edges whose affectation would change the synchronization times of the whole structure.  Finally, in the classification of the networks in Fig. \ref{Fig_6},  we do not see a particular pattern in the number of cycles with three, four, or more nodes, something that in the case of diffusive transport makes the network more resistant to damage \cite{Eraso-metro}.
\\[2mm]
Our findings for the functionality $\langle\mathcal{F}_{\mathrm{s}}(T)\rangle$ show that this quantity gives a global characterization of the effect of damage. However, as it is common in the analysis of complex systems, the ensemble average is just a part of the information and it is necessary to introduce a new quantity to see the effect of damage beyond the dependence on the number of edges. To this end, we explore a measure that considers the probability density $\rho(\tau_0)$ of synchronization times $\tau_0$ and the probability density   $\rho(\tau^\star)$, where
\begin{equation}\label{tau_star}
\tau^\star\equiv\langle\mathcal{F}_{\mathrm{s}}(T)\rangle\tau(T).
\end{equation}
Then, $\tau^\star$ is a scaled version of $\tau(T)$. In this way, the functionality acts as a factor to put $\tau_0$ and $\tau^\star$ in the same scale removing the effect of the number of edges $|\mathcal{E}|$ in the accumulation of damage.
\\[2mm]
In the following, we compare the probability densities $\rho(\tau_0)$ and $\rho(\tau^\star)$. To this end we use 
the Kullback-Leibler divergence, a standard method to calculate the difference between two probability distributions $P(z)$ and $Q(z)$ describing a stochastic variable $z$ \cite{KullbackLiebler_1951,KullbackBook1959}. For continuous distributions, this divergence is given by \cite{KullbackLiebler_1951}
\begin{equation}\label{DKL_C}
    \mathcal{D}_{\mathrm{KL}}[P||Q] = \int P(z)\log\left[\frac{P(z)}{Q(z)}\right]dz.
\end{equation}
Here $Q$ acts as a reference distribution. Also, it is important to emphasize that $\mathcal{D}_{\mathrm{KL}}(P||Q) $ is not a distance in the sense of a metric since the distance between $ P $ and $Q$ is not necessarily the same as between $Q$ and $P$. Also, from the definition in Eq. (\ref{DKL_C}), it is clear that $\mathcal{D}_{\mathrm{KL}}(P||Q)>0$ and is null when $P=Q$.
The Kullback-Leibler divergence is widely used in data science \cite{vandermaaten08a} but also in physical sciences and engineering, for example in the context of turbulence \cite{PhysRevE.97.013107}, the characterization of dynamical systems \cite{PhysRevE.85.031129}, or to compare the activity of vehicles in a transportation system \cite{Metrobus_Sci_Rep_2022}, just to mention a few examples.
\\[2mm]
For the statistical analysis of synchronization times we define   
\begin{equation}
\label{R_KLdist}
\mathcal{R}_{\mathrm{KL}}\{\rho(\tau_0),\rho(\tau^\star)\}
\equiv
\frac{1}{\Delta \tau}\mathcal{D}_{\mathrm{KL}}\left[\rho(\tau_0)\Big|\Big|\frac{\rho(\tau_0)+\rho(\tau^\star)}{2}\right].
\end{equation}
In Eq. (\ref{R_KLdist}), $(\rho(\tau_0)+\rho(\tau^\star))/2$ is the total probability density obtained for the two sets of synchronization times $\tau_0$ and $\tau^\star$. In this manner, the reference distribution $(\rho(\tau_0)+\rho(\tau^\star))/2$ contains total information of the counts of the synchronization times, with this choice we avoid the division by zero in the integral in Eq. (\ref{DKL_C}) used to define $\mathcal{R}_{\mathrm{KL}}\{\rho(\tau_0),\rho(\tau^\star)\}$. $\Delta \tau$ is the bin size in the calculation of $\rho(\tau_0)$ and $\rho(\tau^\star)$. In this manner, $\mathcal{R}_{\mathrm{KL}}$ quantifies the variations between $\rho(\tau_0)$ and the new information associated with the effect of damage. In cases where  $\rho(\tau_0)$ and $\rho(\tau^\star)$ are the same, 
$\mathcal{R}_{\mathrm{KL}}=0$, evidencing that the only effect of damage is the rescaling of synchronization times quantified by 
$\langle\mathcal{F}_{\mathrm{s}}(T)\rangle$. On the other hand, $\mathcal{R}_{\mathrm{KL}}>0$ when  $\rho(\tau_0)$ and $\rho(\tau^\star)$ differ, the value  $\mathcal{R}_{\mathrm{KL}}$ increases with the differences between the two probability densities.
\\[2mm]
\begin{figure*}[t!]
	\begin{center}
		\includegraphics*[width=1.0\textwidth]{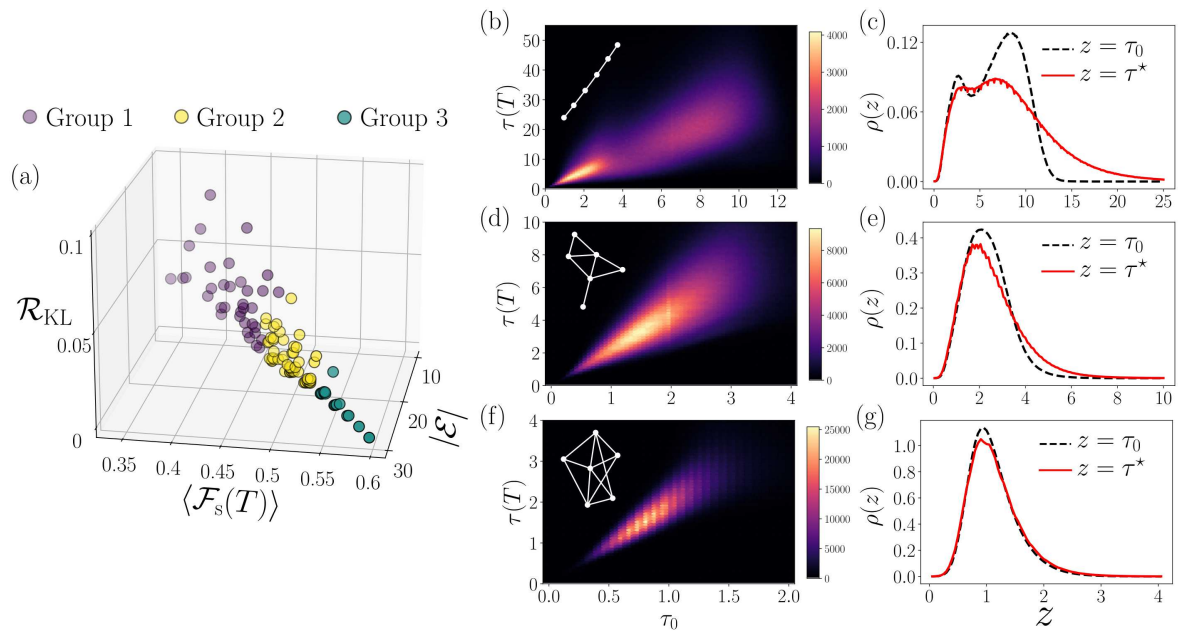}
	\end{center}
	\vspace{-5mm}
	\caption{\label{Fig_7} Comparison of synchronization on networks with $N=6$ under the influence of damage accumulation with $T=100$ and $\alpha=0.5$. (a) Dispersion of values $(|\mathcal{E}|,\,\langle\mathcal{F}_{\mathrm{s}}(T)\rangle,\, \mathcal{R}_{\mathrm{KL}})$ for the 109 networks presented in Fig. \ref{Fig_6}. Numerical values of $\mathcal{R}_{\mathrm{KL}}$ for each graph are obtained using Eq. (\ref{R_KLdist}) to compare the probability density of the times $\rho( \tau^\star)$ of the scaled synchronization time $\tau^\star$ in Eq. (\ref{tau_star}) for the network with damage and the probability density $\rho( \tau_0)$ for the synchronization time on the original structure. (b) Two-dimensional histogram for $10^7$ pairs of values ($\tau_0$, $\tau(T)$) for a linear graph presented as inset, frequency counts are codified in the color bar. (c) Probability densities $\rho(z)$  for synchronization times $z=\tau_0$ and $z=\tau^\star$, we use bins with size $\Delta \tau=0.1$. Panels (d) and (e) show the results for graph number 40 and (f) and (g) for graph 93 in Fig. \ref{Fig_6}. The groups in (a) are obtained using the K-means algorithm (see the main text for a detailed description).}
\end{figure*}
In Fig. \ref{Fig_7} we present the analysis of $\mathcal{R}_{\mathrm{KL}}$ for the 109 graphs in Figs. \ref{Fig_5} and \ref{Fig_6}. We use the information generated with the $10^7$ realizations of times $\tau_0$ and $\tau(T)$ for our model with accumulation of damage with $\alpha=0.5$ and $T=100$. In Fig. \ref{Fig_7}(a) we depict the results as a dispersion of points in three dimensions with coordinates $(|\mathcal{E}|,\,\langle\mathcal{F}_{\mathrm{s}}(T)\rangle,\, \mathcal{R}_{\mathrm{KL}})$. 
We analyze the distribution of 109 points using the K-means algorithm to identify a partition of the values into non-overlapping subgroups (clusters). This unsupervised classification method assigns to data points a cluster such that the sum of the squared distance between the data points and the cluster’s centroid (arithmetic mean of all the data points that belong to that cluster) is at the minimum \cite{Lloyd_82}, the number of clusters $K$ has to be predetermined.
We use the Python library {\it Scikit-learn} \cite{scikit-learn} for an implementation of the K-means algorithm.  In Fig. \ref{Fig_7}(a) we show the results obtained for $K=3$, we refer to each set of points as groups 1 to 3. The points in each group are presented with different colors. Here it is worth mentioning that other values of $K$ can be explored; however, we center our discussion using three groups. This classification for the data shows graphs in group 1 are affected significantly by damage, having low functionality, and with major changes in the probability densities $\rho(\tau_0)$ and $\rho(\tau^\star)$. In group 2 the $\langle\mathcal{F}_{\mathrm{s}}(T)\rangle$ increase in comparison to the values in group 1  and the $\mathcal{R}_{\mathrm{KL}}$ are reduced, revealing that the probabilities densities of synchronization times present moderate variations. Finally, in group 3, the graphs are more resistant to damage, and $\rho(\tau_0)$ and $\rho(\tau^\star)$ suffer minimal variations. To illustrate the features found in the classification, in Figs. \ref{Fig_7}(b)-(g), we discuss the statistical analysis of the synchronization times in three particular graphs, each in one of the groups 1 to 3.
\\[2mm]
In Figs.  \ref{Fig_7}(b)-(c), the linear graph (graph 6 in Fig. \ref{Fig_6}) is analyzed. The synchronization in this structure is classified in group 1. In panel (b) we show the two-dimensional histogram with the counts of $10^7$ pairs $(\tau_0,\,\tau(T))$, the bin counts are codified in the color bar and the respective network is presented as an inset. The results in this representation show the variations of $\tau(T)$ due to the damage and in general, the relation with $\tau_0$ is nonlinear. In this case, $\langle\mathcal{F}_{\mathrm{s}}(T)\rangle$ only contains partial information of the relation between $\tau_0$ and $\tau(T)$. Additional modifications associated with the accumulation of damage in the synchronization are observed in panel (c) with the statistical analysis of $\tau_0$ and the scaled time $\tau^\star$.  Here, it is worth noticing that $\rho(\tau_0)$ is bimodal, revealing that in the linear graph, some initial conditions generate in high proportion the synchronization with times around $\tau_0=2.5$ whereas, from other initial conditions, the system synchronizes around $\tau_0=9$. The effect of damage reduces the two relative maximums observed in $\rho(\tau_0)$. The probability densities differ significantly producing higher values of  $\mathcal{R}_{\mathrm{KL}}$ in Eq. (\ref{R_KLdist}). Similar features are found in all the graphs in group 1. 
\\[2mm]
Similarly, in Figs. \ref{Fig_7}(d)-(e), we present the analysis of the graph 40 in Fig. \ref{Fig_6} classified in group 2. In panel (d), we see that the pairs $(\tau_0,\,\tau(T))$ disperse around a line. The analysis of the probability densities $\rho(\tau_0)$ and $\rho(\tau^\star)$ in (e) show moderate differences between the two curves evidenced with an intermediate value of  $\mathcal{R}_{\mathrm{KL}}$, these characteristics are present in all the graphs in group 2. Finally, we have group 3 in the classification. The effect of damage in this group is illustrated with the analysis of graph 93 in Fig. \ref{Fig_6} presented in Figs. \ref{Fig_7}(f) and (g). The results show that the effect of damage is well described by a linear relation, a fact also evidenced in the minimal differences between $\rho(\tau_0)$ and $\rho(\tau^\star)$ producing a $\mathcal{R}_{\mathrm{KL}}$ closer to zero.
\\[2mm]
The results in Fig. \ref{Fig_7} show that $\mathcal{R}_{\mathrm{KL}}$ captures the effect of the accumulation of damage in synchronization processes beyond the number of edges. As a final test for this quantity, we evaluate the functionality $\langle\mathcal{F}_{\mathrm{s}}(T)\rangle$ and  $\mathcal{R}_{\mathrm{KL}}$ for the four networks with $N=100$ nodes analyzed in Fig. \ref{Fig_2}. Our findings are presented in Table \ref{Tab_1}, where we analyze $10^5$ pairs of values generated for $T=1000$, $\alpha=0.5$ considering synchronization times for $r=0.99$. The values of $\mathcal{R}_{\mathrm{KL}}$ are calculated using Eq. (\ref{R_KLdist}) with probability densities obtained with bin counts with $\Delta \tau=0.1$, the probability densities for $\rho(\tau_0)$ were analyzed in Fig. \ref{Fig_2}.
\begin{table}[t]
	\centering
	\begin{tabular}{c c  c  c }
		\hline    
		{\bf Network} &  $|\mathcal{E}|$ &  $\langle\mathcal{F}_{\mathrm{s}}(T)\rangle$  & $\mathcal{R}_{\mathrm{KL}}$\\[0.5mm]
		\hline
		Knight     &     576             & 0.7072   &   0.0138\\
		Barab\'asi-Albert            &     392       &        0.6185   &   0.0208\\
		Erd\H{o}s-R\'eny    & 392     &  0.593   &   0.1117\\
		Wheel             & 396     &  0.5907   &   0.1532\\           
 \hline
	\end{tabular}
	\caption{\label{Tab_1} Characterization of synchronization and accumulation of damage for the networks with $N=100$ nodes   in Fig. \ref{Fig_2}. The values are obtained from the statistical analysis of $10^5$ pairs of synchronization times $(\tau_0,\, \tau(T))$ generated for $T=1000$, $\alpha=0.5$ and $r=0.99$.}
\end{table}
\\[2mm]
The values reported in Table \ref{Tab_1} allow us to classify the networks studied in Fig. \ref{Fig_2} in terms of their resistance to damage generated with the preferential attachment algorithm and the modifications in the synchronization times. The networks are sorted according to the values $\langle\mathcal{F}_{\mathrm{s}}(T)\rangle$ in decreasing order presenting the networks from the most robust to the most fragile. The results show that the knight network is the most resistant with the highest functionality, a fact attributable to the number of edges. Then, we have the Barab\'asi-Albert network with a lower number of edges. In both most resistant networks, we observe small values of $\mathcal{R}_{\mathrm{KL}}$ showing that in these cases the damage only rescales synchronization times but the probability densities of the scaled time $\tau^\star$ are similar to the found in Figs. \ref{Fig_2}(a) and (d) for $\rho(\tau_0)$. This behavior is analogous to the observed in group 3 for synchronization on graphs with $N=6$. In contrast, with the same number of edges as the Barab\'asi-Albert network, the Erd\H{o}s-R\'eny network has lower functionality but the effect of damage is more clear with the value of $\mathcal{R}_{\mathrm{KL}}$ that increases. A similar result is found in the wheel. In fact, the Erd\H{o}s-R\'eny network and the wheel suffer intermediate modifications in the probability densities of the synchronization times; the characteristics of the changes are similar to those observed in group 2 in the analysis in Fig. \ref{Fig_7}.
\section{Conclusions}
In this paper, we explored the consequences of the accumulation of damage in the synchronization process of networks of identical Kuramoto oscillators. The algorithm implemented generates random cumulative damage in the links producing a directed weighted network modeling the detrimental of the couplings between oscillators. In this system, cumulative damage generates changes in the synchronization times. We propose global measures to quantify the changes in the synchronization process as the damage increases in the network.  The first one that we call functionality of synchronization $\mathcal{F}_\mathrm{s}(T)$ is defined as the quotient between the synchronization time when the network does not have damage and the synchronization time when the network is affected by the damage, in both cases we consider the same initial conditions of the oscillators as random phases. The values of $\mathcal{F}_\mathrm{s}(T)$ decrease with damage and capture global features of the reduction of the capacity of the networks to reach synchronization; however, when we analyze the ensemble average of this quantity most of the captured information relates to how the number of links in the graph improves its response to damage. The second measure $\mathcal{R}_{\mathrm{KL}}$ captures the changes that the probability densities of the synchronization times experiment with damage, its value increases when the changes in the probabilities are more noticeable. We apply all those ideas to study the synchronization process in different topologies that admit fully synchronization for most of the initial conditions as a wheel, a knight graph, an Erd\H{o}s-R\'eny network, and a Barab\'asi-Albert network. Also, we analyze the set of non-isomorphic connected graphs with six nodes.
\\[2mm]
The findings of our research show that
globally the cumulative random damage re-scales synchronization times of the networks, this effect is measured by means of $\mathcal{F}_\mathrm{s}(T)$ that depends on a non-linear way of the number of edges of the networks,  its ensemble average $\langle\mathcal{F}_\mathrm{s}(T)\rangle$ is able to quantify some differences in the synchronization process of networks with damage beyond the number of edges and due to their topology, nevertheless the deviations are very small and it is necessary many realizations of the process to be able to capture them. On the other hand, the measure $\mathcal{R}_{\mathrm{KL}}$ uses more information about the synchronization process and evaluates better the effects of cumulative damage. The combination of both measures  $\langle\mathcal{F}_\mathrm{s}(T)\rangle$ and  $\mathcal{R}_{\mathrm{KL}}$ results in a classification of networks that synchronize in three groups according to whether their damage tolerance is low, medium or high focusing in their topology rather than their number of edges as the case of the set of non-isomorphic connected graphs with six nodes.
\\[2mm]
The methods and results of this work contribute to a better understanding of the synchronization, in general, they represent an approach to understanding the degradation of the functions of complex systems that accumulate damage in time. As a generalization of the model, it would be interesting to consider the process of accumulation of damage commensurable with the synchronization of the system. This problem can be studied in the framework of time-varying coupling networks presented in  \cite{GHOSH20221,Zhou_Zou_Guan_Liu_Boccaletti_2016, delGenio_Romance_Criado_Boccaletti_2015, Boccaletti_Hwang_Chavez_Amann_Kurths_Pecora_2006,Porfiri} including the analysis of changes of the stability of the synchronization process in networks with damage. Furthermore, since  we are studying a model with identical oscillators, new research can include features such as heterogeneous oscillators or the presence of noise.   The approach introduced paves the way for the exploration of other dynamical systems in the presence of damage.

\section{Appendix}
\subsection{Effect of damage on the linear approximation of the Kuramoto model}
\label{Append1}
In the Kuramoto model, it has been shown that for initial conditions closer to the synchronization is valid a linear approximation; in this particular case,
synchronization times scale with the inverse of the second non-zero smallest eigenvalue of the Laplacian matrix that defines the connectivity of the network  \cite{Almendral_2007,Grabow_Hill_Grosskinsky_Timme_2010,Grabow_Grosskinsky_Timme_2011}. Also, under this linear approximation of the Kuramoto model of identical oscillators, the stability of the synchronous state can be assessed using a  master stability function approach  \cite{Barahona_Pecora_2002,Pecora1998} that   gives the necessary conditions for the system to synchronize and establishes restrictions on  the eigenvalues of the Laplacian matrix of the network that guarantee the negativity of the maximum Lyapunov exponent
\cite{Barahona_Pecora_2002,Pecora1998, PhysRevLett.96.114102, ARENAS200627}.
\\[2mm]
In this appendix, we present the modifications that introduce the effect of damage in the linear approximation of the Kuramoto model. For small values of $\theta_j(t)-\theta_i(t)$, is valid the linear approximation of Eq. (\ref{Kuramoto})
\begin{equation}
    \frac{d\theta_i(t)}{dt}\approx\sum_{j=1}^{N
    }\Omega_{ij}(T)[\theta_j(t)-\theta_i(t)]=-\sum_{j=1}^N (\mathcal{S}_i(T)\delta_{ij}-\Omega_{ij}(T))\theta_j(t),
\label{linear_approx1}
\end{equation}
where the strength  $\mathcal{S}_i(T)$ of the node $i$ is defined as  $\mathcal{S}_i(T)\equiv\sum_{\ell=1}^N \Omega_{i\ell}(T)$ and $\delta_{ij}$ denotes the Kronecker delta. In this manner, considering the form of the Laplacian matrix of a weighted network, we define the elements $L_{ij}(T)$ of the Laplacian matrix $\mathbf{L}(T)$ as
\begin{equation}
    L_{ij}(T)\equiv\mathcal{S}_i(T)\delta_{ij}-\Omega_{ij}(T).
\end{equation}
%
Therefore, the linear approximation in Eq. (\ref{linear_approx1}) defines the dynamical process
\begin{equation}\label{dtheta_linear}
\frac{d\theta_i(t)}{dt}=-\sum_{j=1}^N L_{ij}(T)\theta_j(t).
\end{equation}
The integration of Eq. (\ref{dtheta_linear}) leads to
\begin{equation}\label{linear_exp}
\theta_i(t)=\sum_{j=1}^N\left(e^{-t \mathbf{L}(T)}\right)_{ij}\theta_j(0).
\end{equation}
Using Dirac's notation for the eigenvectors, we have a set of right eigenvectors $\{\left|\Psi_j(T)\right\rangle\}_{j=1}^N$  that satisfy the eigenvalue equation
$\mathbf{L}(T)\left|\Psi_j(T)\right\rangle=\mu_j(T)\left|\Psi_j(T)\right\rangle$ for $j=1,\ldots,N$. With this information, we define the matrix $\mathbf{Q}(T)$  with elements $Q(T)_{ij}=\left\langle i|\Psi_j(T)\right\rangle$  and the diagonal matrix $\mathbf{\Lambda}(t,T)=\textrm{diag}(e^{-t\mu_1(T)},e^{-t\mu_2(T)},\ldots,e^{-t\mu_N(T)})$. These matrices satisfy
\begin{equation}
e^{-t \mathbf{L}(T)}=\mathbf{Q}(T)\mathbf{\Lambda}(t,T)\mathbf{Q}(T)^{-1},
\end{equation}
where $\mathbf{Q}(T)^{-1}$ is the inverse of $\mathbf{Q}(T)$. Using the matrix $\mathbf{Q}(T)^{-1}$, we define the set of left eigenvectors $\{\left\langle \bar{\Psi}_i(T)\right|\}_{i=1}^N$ with components $\left\langle \bar{\Psi}_i(T)|j\right\rangle=(\mathbf{Q}(T)^{-1})_{ij}$. Therefore, the solution for the linear dynamics in  Eq. (\ref{linear_exp}) takes the form
\begin{equation}\label{theta_i_eigensys}
    \theta_i(t)=\sum_{j=1}^N\sum_{\ell=1}^N e^{-t\mu_\ell(T)}\langle i|\Psi_\ell(T)\rangle \langle \bar{\Psi}_\ell(T)|j\rangle \theta_j(0).
\end{equation}
Then, the effect of damage accumulation in the linear approximation defines a process that can be explored analytically using the same methods developed in our study of the Kuramoto dynamics; however, all the temporal evolution is determined by the eigenvalues and eigenvectors of $\mathbf{L}(T)$. Here it is worth to notice that  the temporal evolution on heterogeneous weighted networks, i.e., in cases where the strength of nodes  $\mathcal{S}_i(T)$ is not a constant, the linear dynamics  is not equivalent  
to a process of diffusive transport associated to continuous-time random walks defined by the normalized Laplacian with elements $    \mathcal{L}_{ij}(T)\equiv\delta_{ij}-\Omega_{ij}(T)/\mathcal{S}_i(T)$ [see Ref. \cite{Eraso_Hernandez_2021} for a detailed discussion of the discrete-time random walk dynamics on networks with accumulation of damage with transition probabilities between nodes $i$, $j$ given by $w_{i\to j}(T)=\Omega_{ij}(T)/\mathcal{S}_i(T)$].
\\[2mm]
\begin{figure}[t]
    \begin{center}
   	 \includegraphics*[width=1.0\textwidth]{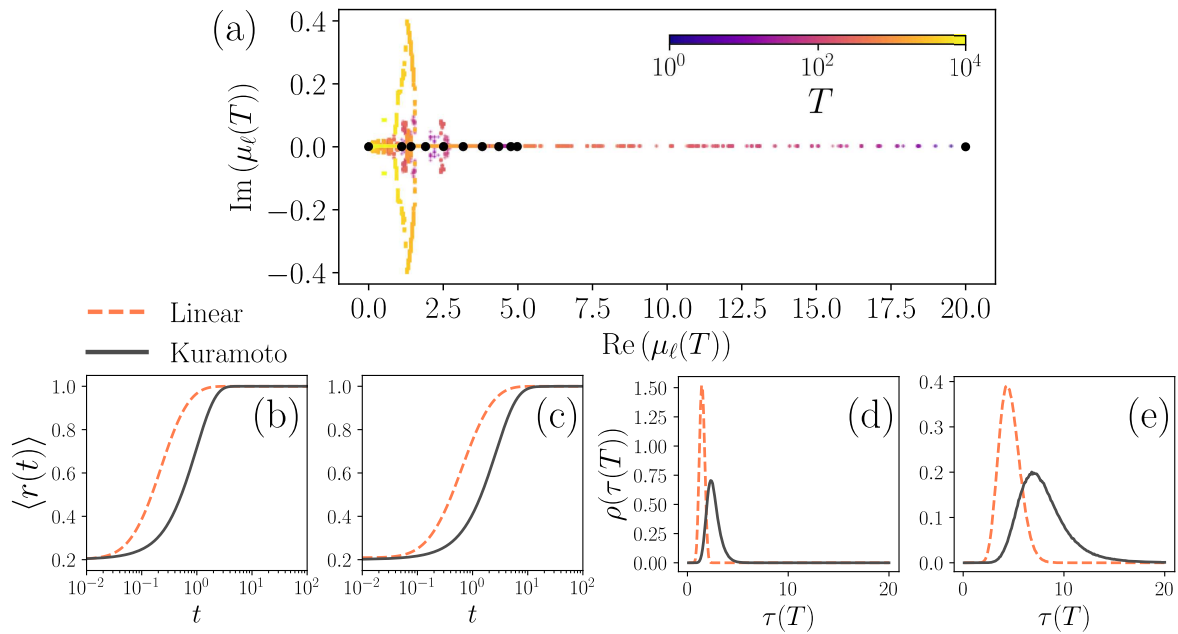}
    \end{center}
    \vspace{-5mm}
    \caption{\label{Fig_8} Effect of damage with $\alpha=0.5$ for the linear approximation and the Kuramoto dynamics on a wheel with $N=20$ nodes. (a) Set of eigenvalues $\{\mu_\ell(T)\}_{\ell=1}^N$ of $\mathbf{L}(T)$ in the complex plane for $T=1,\ldots,10^4$ codified in the colorbar. We represent with black dots the eigenvalues of the Laplacian matrix without damage $\mathbf{L}(0)$.  (b) Evolution of the ensemble average of the order parameter $\langle r(t)\rangle$ at $T=0$ for the network using the linear evolution and the Kuramoto model,  average values are obtained considering $1000$ realizations. (c) Results obtained for $\langle r(t)\rangle$ for the system with damage at $T=1000$. (d) Probability densities $\rho(\tau(T))$ of synchronization times $\tau(T)$, the results are generated using $10^6$ realizations of the linear and Kuramoto models to obtain the time $\tau(T)$ when the order parameter $r(t)$ in Eq. (\ref{orderparam}) reaches the value $r=0.99$ for $T=0$, the same analysis is presented in (e) for $T=1000$, bin sizes $\Delta \tau=0.1$ are used. In panels (b)-(d), dashed lines indicate the results for the linear model obtained through numerical evaluation of Eq. (\ref{theta_i_eigensys}), continuous lines show the results using the numerical integration of the Kuramoto model in Eq. (\ref{Kuramoto}),  both dynamics used random initial conditions uniformly distributed in the interval $[0,2\pi)$.  }   
\end{figure}
To illustrate characteristic features introduced by the effect of damage in the linear dynamics, in Fig. \ref{Fig_8}, we explore the algorithm for damage accumulation with $\alpha=0.5$ on the wheel graph with $N=20$ discussed in Fig. \ref{Fig_3}. In Fig. \ref{Fig_8}(a), we depict the eigenvalues $\{\mu_\ell(T) \}_{\ell=1}^N$ of the Laplacian matrix $\mathbf{L}(T)$ for one realization of the algorithm with $T=1,\ldots ,10^4$. The sets of eigenvalues are presented in the complex plane with different colors for each $T$, the eigenvalues for the structure without damage ($T=0$) are shown with black dots. In this representation of the results is evident how the asymmetry of $\mathbf{\Omega}(T)$ produces complex values of $\mu_\ell(T)$.
\\[2mm]
On the other hand, in panels \ref{Fig_8}(b)-(c), we compare the ensemble average $\langle r(t)\rangle$ as a function of $t$ for the Kuramoto model in Eq. (\ref{Kuramoto}) and the linear approximation obtained from the numerical evaluation of the analytical result in Eq. (\ref{theta_i_eigensys}). The values are obtained for the case without damage $T=0$ (panel (b)) and $T=1000$ (panel (c)). Using a similar approach, in Figs. \ref{Fig_8}(d)-(e), we present the statistical analysis of the time $\tau(T)$ necessary to reach for the first time the order parameter $r=0.99$ for $T=0$ (panel (d)) and $T=1000$ (panel (e)). In all the realizations in Figs. \ref{Fig_8}(b)-(d) we consider random initial conditions for the phases $\theta_i(0)$, with values chosen  random (uniformly distributed) in the interval $[0,2\pi)$.
\\[2mm]
The results in Fig. \ref{Fig_8} reveal the marked differences between the linear approximation and the Kuramoto model. On average, synchronization times for the linear case are lower than the values observed in the non-linear dynamics. The discrepancies are due to the use of initial random phases for which the approximation in Eq. (\ref{linear_approx1}) is not valid. However, in other types of problems with initial conditions closer to synchronized phases, the linear dynamics with phases in Eq. (\ref{theta_i_eigensys}) gives a good approximation for the synchronization where we can see directly that the eigenvalues of the Laplacian matrix are associated with characteristic times of the process. In our general modeling, we are interested in a functionality measure defined in terms of the capacity of a system to reach synchronization from random initial phases. To this end,  in the discussion presented in the main text, we restrict our study to the Kuramoto model.
\subsection{Variations of the functionality with the threshold value $r$}
\label{Append2}
\begin{figure}[t]
    \begin{center}
   	 \includegraphics*[width=1.0\textwidth]{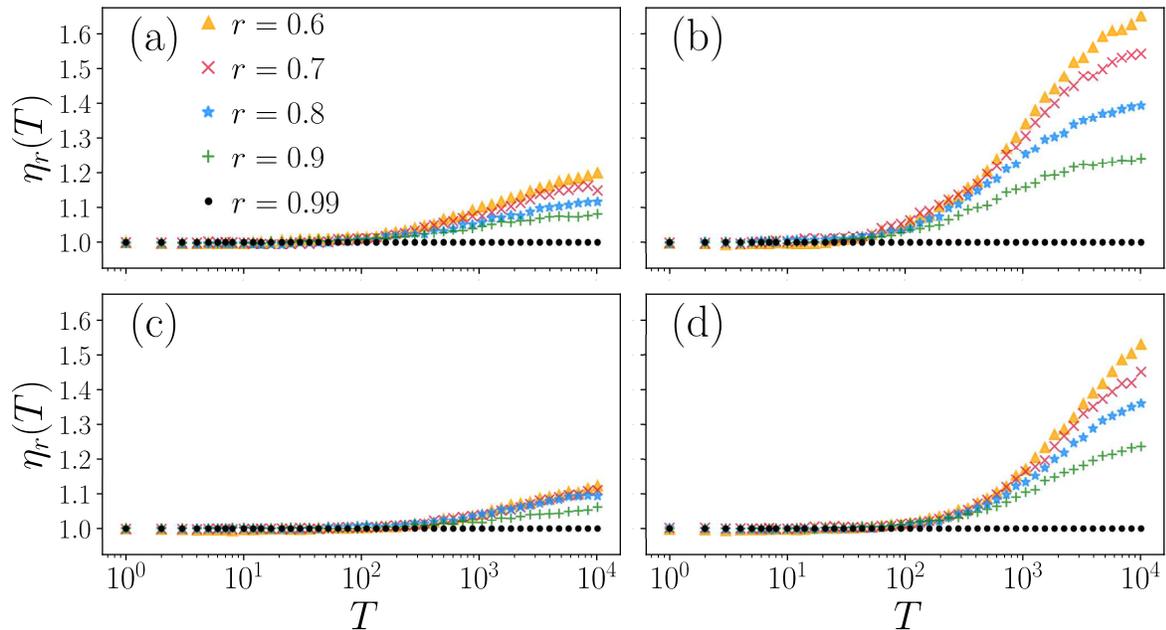}
    \end{center}
    \vspace{-5mm}
    \caption{\label{Fig_9} $\eta_r(T)$ as a function of $T$  using the threshold $r$ for a wheel graph and a Barab\'asi-Albert network with $N=100$. The ensemble averages in $\eta_r(T)$ defined by Eq. (\ref{eta_def}) are generated using 1000 realizations for $\langle\mathcal{F}_\mathrm{s}(T)\rangle_r$ with threshold values $r=0.6,0.7,0.8,0.9$ and compared with the result obtained for $r=0.99$ for $1\leq T\leq 10^4$. Wheel graph with (a)  $\alpha=0.5$ and (b) $\alpha=1.0$. Barab\'asi-Albert network with (c) $\alpha=0.5$ and (d) $\alpha=1.0$.}
    \end{figure}
    
In our study of the effects of damage in synchronization in the main text, we selected a particular threshold value of the order parameter $r$ in Eq. (\ref{orderparam}). It was convenient to choose the threshold $r=0.99$ to study configurations close to complete synchronization (obtained when $r=1$). Nevertheless, it is important to understand the effects of choosing different values of $r$ and how the threshold selected can affect our findings. In this appendix, we explore numerically the implications of the threshold value $r$ in the results of the ensemble average of the functionality $\mathcal{F}_{\mathrm{s}}(T)$ defined in Eq. (\ref{Fun}). We explore two networks for two values of the parameter $\alpha$ and $1\leq T\leq 10^4$.
\\[2mm]
We define ratio
\begin{equation}\label{eta_def}
	\eta_r(T)\equiv\frac{\langle\mathcal{F}_\mathrm{s}(T)\rangle_r}{\langle\mathcal{F}_\mathrm{s}(T)\rangle_{r=0.99}},
\end{equation}
where $\langle\mathcal{F}_\mathrm{s}(T)\rangle_r$ denotes the ensemble average of functionality for a given threshold value $r$; in particular $\langle\mathcal{F}_\mathrm{s}(T)\rangle_{r=0.99}$ is obtained with $r=0.99$, the value we used in the different analyses discussed in the main text. In Fig. \ref{Fig_9}, we present numerical results obtained for $\eta_r(T)$ using Monte Carlo simulations, the methods implemented are the same as described for Figs. \ref{Fig_2} and \ref{Fig_3} but now using the threshold $r=0.6, 0.7, 0.8, 0.9$ and $0.99$ to define the synchronization times and $\alpha=0.5,\,1$ for the wheel graph and the  Barab\'asi-Albert network with size $N=100$ studied in  Fig. \ref{Fig_2}, the ensemble averages were obtained using 1000 realizations.
\\[2mm]
In the panels in Fig. \ref{Fig_9}, we plot $\eta_r(T)$ as a function of $T$, each distribution of points is obtained with a particular value of $r$ codified in the markers in panel (a).  The dots depict the case $\eta_r(T)=1$ when $\langle\mathcal{F}_\mathrm{s}(T)\rangle_r$  coincides $\langle\mathcal{F}_\mathrm{s}(T)\rangle_{r=0.99}$.  In Figs. \ref{Fig_9}(a)-(b) we depict the results for the wheel graph using $\alpha=0.5$ and $\alpha=1.0$, respectively.  Similarly in Figs. \ref{Fig_9}(c)-(d), we present the results for the Barab\'asi-Albert network. In both cases, the numerical values reveal that for $\alpha=0.50$, $\eta_r(T)$ are close 1 showing to $\langle\mathcal{F}_\mathrm{s}(T)\rangle_r$ are similar for the different values of $r$ to  the result obtained for $r=0.99$.   This is evident in both the structures, the wheel and the Barab\'asi-Albert network. However, in panels (b) and (d) generated for $\alpha=1$,  the changes of  $\langle\mathcal{F}_\mathrm{s}(T)\rangle_r$ are more noticeable evidenced by higher values of $\eta_r(T)$ for $T\geq 10^3$.  The results in Fig. \ref{Fig_9} characterize the effect of the threshold $r$ and how  affects the values of   $\langle\mathcal{F}_\mathrm{s}(T)\rangle$ and show that for $r$ close to 1.0, maintaining $\alpha<1$, the $\langle\mathcal{F}_\mathrm{s}(T)\rangle$ are not significantly affected by the choice of $r$.

\section*{Acknowledgments}
LKEH acknowledges support from CONACYT M\'exico. APR acknowledges financial support from Ciencia de Frontera 2019 (CONACYT), grant 10872. APR acknowledges fruitful discussions with Thomas Michelitsch and Jicun Wang Michelitsch on aging processes in complex systems.

\section*{References}

\providecommand{\noopsort}[1]{}\providecommand{\singleletter}[1]{#1}%
\providecommand{\newblock}{}

\end{document}